\documentclass{ws-procs9x6}

\begin{document}

\title{Power-like abundance of elements in Universe}

\author{G.Wilk}

\address{The Andrzej So\l tan Institute for Nuclear Studies; 
Ho\.za 69; 00-689 Warsaw, Poland\\E-mail: wilk@fuw.edu.pl}

\author{Z.W\L odarczyk}

\address{Institute of Physics, \'Swi\c{e}tokrzyska Academy;
Konopnickiej 15; 25-405 Kielce, Poland\\
E-mail: wlod@pu.kielce.pl}

\maketitle

\abstracts{
The apparently observed power-like abundance of elements in Universe
is discussed in more detail with special emphasis put on the
strangelets. 
}  
\maketitle

\section{Introduction}

We would like to bring ones attention to the apparently not so well
know fact that the abundance of all known elements in the Universe
follow the characteristic (and so far unexplained) power law
dependence as is clearly demonstrated in Fig. 1 \footnote{Consecutive
steps of histogram in Fig. 1 denote the following nuclei: Ne, (Mg,
Si), S, (K, Ca), Fe, (Cu, Zn), (Kr, Sr, Zr), (Te, Xe, Ba), (Rare
earths), (Os, Ir,Pt, Pb).}. Between the heaviest atomic elements and
neutron stars (which can also be regarded as a kind of "element")
lies then a vast unpopulated "nuclear desert". We would like to
propose that actually this "nuclear desert" can be populated by
nuggets of the Strange Quark Matter (QSM) called {\it strangelets}
which, according to our estimations \cite{WW,APP}, follow the
$A^{-7.5}$ power-like dependence indicated in Fig. 1. Referring to
our works \cite{WW,APP} for details we would like to mention only
here that all known candidates for strangelets \cite{S} fit very
nicely to the expected behaviour of such objects (cf. Fig. 4 in
\cite{APP}). In order to solve the apparent contradiction that,
although strangelets seem to follow the normal nuclear $A$-dependence
of sizes, i.e., $r_{S} = r_0 A^{1/3}$ (with $r_0\sim 1.2$ fm
\cite{WW}), nevertheless they do occur deeply in the atmosphere, we
have proposed \cite{WW} the following scenario: strangelets entering
the atmosphere of Earth are big, with atomic number $A$ of the order
of $A = 1000\div 2000$. Being rather "big bags" of quarks than normal
nuclei they do not, however, disintegrate during the consecutive
interactions with the atmospheric nuclei but simply become
smaller. Only after reaching some critical size of $A_{crit} =
200\div 400$ they do disintegrate into baryons\footnote{It is quite
plausible that the famous {\it Centauro} events are caused by
precisely such mechanism, see \cite{Centauro} for details.}.   
\vspace{-1cm}
\begin{figure}[h]
\begin{minipage}[t]{0.475\linewidth}
\centering
\includegraphics[height=5.cm,width=6.cm]{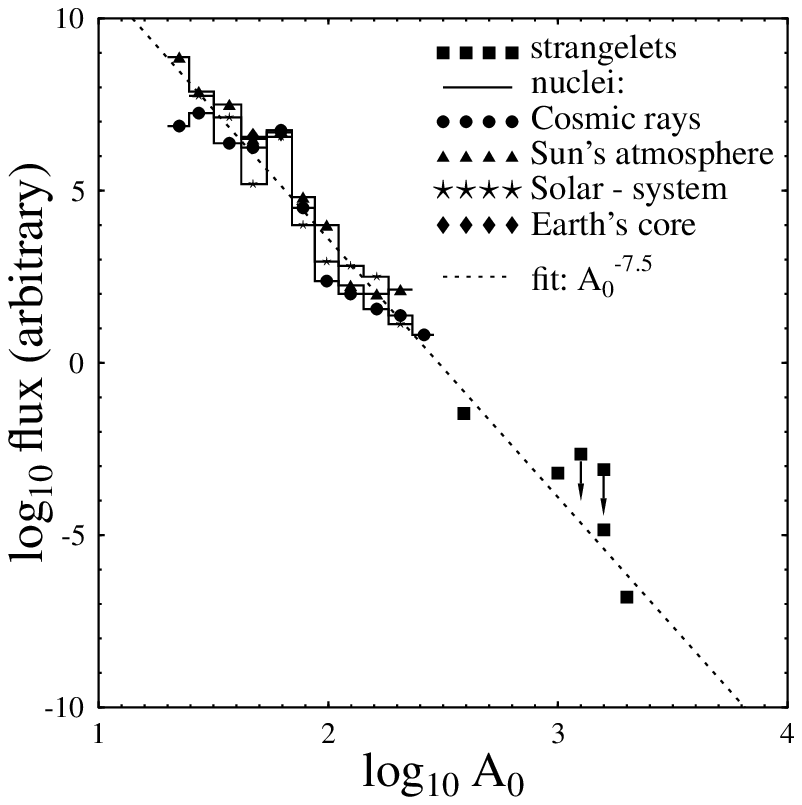}
\vspace{-10mm}
\caption{Comparison of the estimated mass spectrum $N(A_{0})$ for
strangelets with the known abundance of elements in the Universe
\protect\cite{Zhdanov}. } 
\end{minipage}\hfill
\begin{minipage}[t]{0.475\linewidth}
\centering
\includegraphics[height=5.cm,width=6.cm]{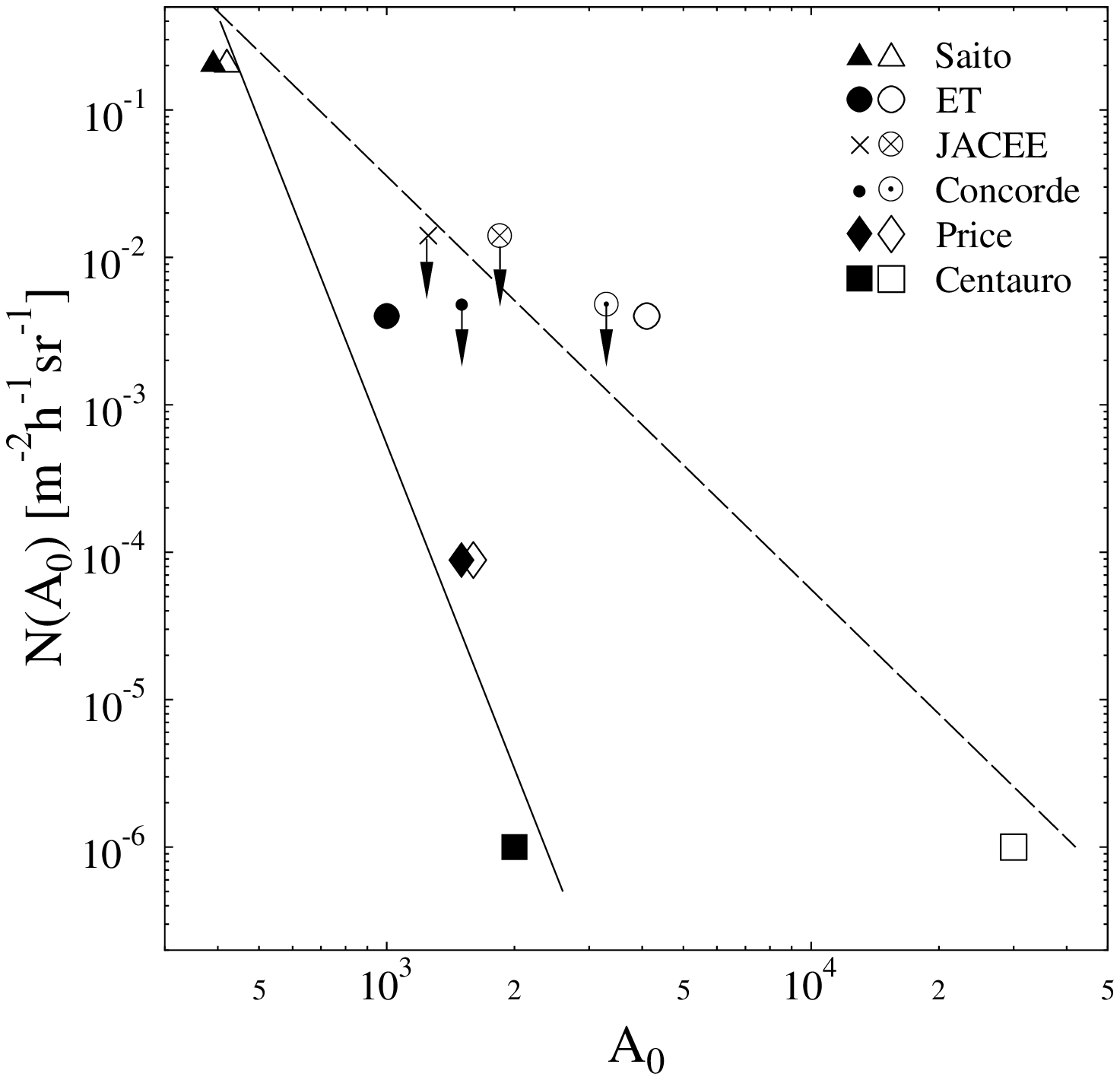}
\vspace{-10mm}
\caption{The estimation of the expected flux of strangelets on the
border of atmosphere, $N(A_{0})$, as a function of their mass number.
See \protect\cite{WW} for further details.
} 
\end{minipage}
\end{figure}
\vspace{-1.cm}
\begin{figure}[h]
\begin{minipage}[t]{0.475\linewidth}
\centering
\includegraphics[height=5.cm,width=6.cm]{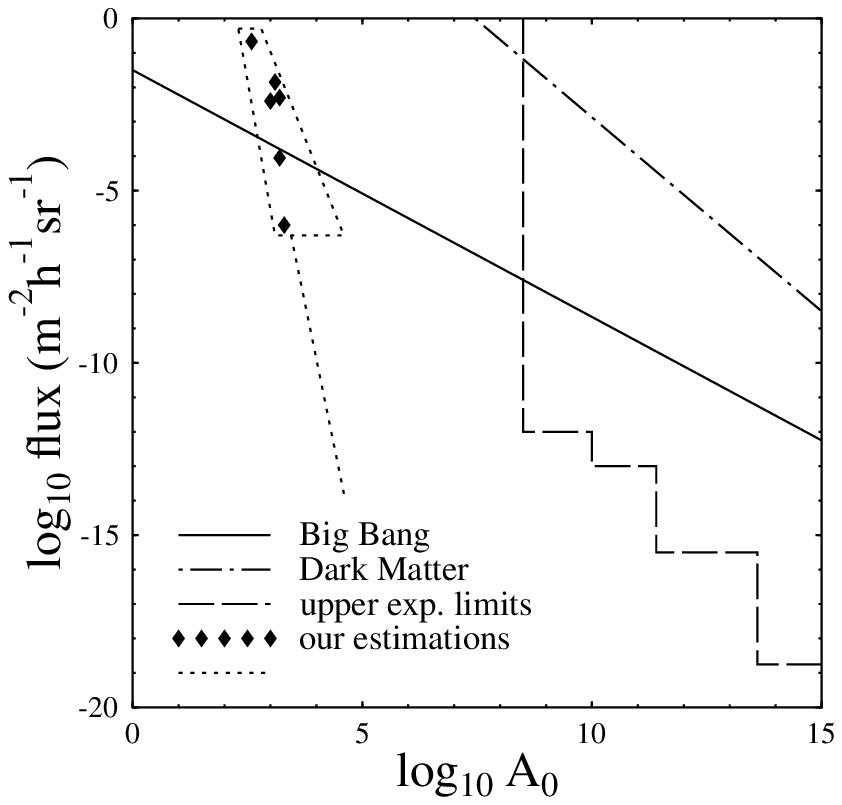}
\vspace{-10mm}
\caption{The expected flux (our results \protect\cite{APP}) of strangelets
compared with the upper experimental limits, compiled by Price
\protect\cite{Price}, and predicted astrophysical limits.}
\end{minipage}\hfill
\begin{minipage}[t]{0.475\linewidth}
\centering
\includegraphics[height=5.cm,width=6.cm]{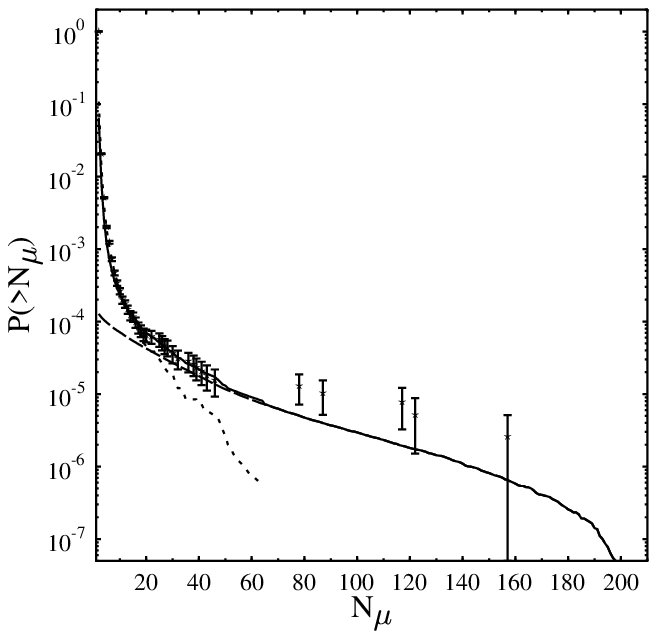}
\vspace{-10mm}
\caption{Integral multiplicity distribution of muons for the
CosmoLEP data (stars) compared with contributions from "normal"
composition of primary nuclei (dotted line), with primary strangelets
of $A=400$ (broken line) and their sum (full line)\protect\cite{Ryb}.}    
\end{minipage}
\end{figure}

Such scenario allows to estimate the flux of strangelets reaching our
atmosphere, cf. Fig. 2, and to compare it with some other
cosmological estimations, see Fig. 3 \footnote{In Fig. 2 the
estimations are provided for two specific realizations of our
propositions \cite{WW} out of which only one (solid line) results in
power-like fit $\sim A_0^{-7.5}$ shown in Figs. 1 and 3.}. 
We would
like to stress here that with the flux of strangelets estimated in
this way we were able to fit data on distribution of muons obtained
by ALEPH detector at the CosmoLEP experiment (see Fig. 4 and
\cite{Ryb} for details) (and it should be stressed that only in our
approach can one fit the last $5$ events in Fig. 5 with unexpectedly
large multiplicities $N_{\mu}$ (up to 150) which cannot be explained,
even assuming pure iron primaries)\footnote{To do so we have used the
so-called "normal" chemical composition of primaries with 40\%
protons, 20\% helium, 20\% CNO mixture, 10\% Ne-S mixture, and 10\%
Fe. It can describe low multiplicity ($N_{\mu} \leq 20$) region only.
Muon multiplicity from strangelet induced showers are very broad. As
can be seen, the small amount of strangelets (with the smallest
possible mass number $A=400$ (the critical mass to be $A_{crit}=320$
here)) in the primary flux can accommodate experimental data. Taking
into account the registration efficiency for different types of
primaries one can estimate the amount of strangelets in the primary
cosmic flux. In order to describe the observed rate of high
multiplicity events one needs the relative flux of strangelets
$F_S/F_{tot} \simeq 2.4 \cdot 10^{-5}$ (at the same energy per
particle). 
}.
The above results clearly demonstrated the plausibility of our
supposition expressed in Fig. 1, namely that it is quite possible
that the power-law found when discussing the abundance of known
elements in the Universe \cite{Zhdanov} is followed by the nuggets of
SQM, strangelets, signals of which are accumulating
\cite{S,Centauro,APP}. The $A^{-7.5}$ dependence found (and in
general, the reason for such power law to show up) is, to our
knowledge, yet unexplained. We would like to mention here a very
simple reasoning leading to such behaviour. Let us consider an object
(nucleus or strangelet) carrying mass number $A_i$, which absorbs
neutrons with probability $W=\Phi \sigma = A_i v \sigma$. In time $t$
it travels distance $L=vt$. If absorptions of neutrons proceeds with
cross section $\sigma$ and density of neutrons is $n$, then the
increment of mass is $\partial A_i = nA_iL\sigma\partial t/t$. For
the $i$-th such object
\begin{equation}
\frac{\partial A_i}{A_i}\, =\, \alpha\cdot \frac{\partial t}{t}\quad
\Longrightarrow \quad A_i(t) = m
\left(\frac{t}{t_i}\right)^{\alpha}\quad {\rm where}\quad \alpha =
nL\sigma \, 
\end{equation}
(where initial condition that $i$-th object occurs in time $t_i$ with
mass $A_i(t_i)=m$ was used). Probability of forming an object of mass
$A_i(t) < A$ is then
\begin{equation}
P(A_i(t)<A)\, =\, P\left(
t_i > \frac{m^{1/\alpha}\cdot t}{A^{1/\alpha}}
                   \right)\, . \label{eq:PPP}
\end{equation}
Assuming now uniform distribution of occuring of such objects in the
system, the probability of adding such object in the unit time
interval to the system is equal to $P(t_i) = 1/t$ what leads to
\begin{equation}
P\left(
        t_i > \frac{m^{1/\alpha}\cdot t}{A^{1/\alpha}}
                   \right)\, =\, 
 1\, -\, P\left(
        t_i \leq \frac{m^{1/\alpha}\cdot t}{A^{1/\alpha}}
                   \right)\, =\, 1\, -\,
\frac{m^{1/\alpha}}{A^{1/\alpha}}\, .
\end{equation}
Using now (\ref{eq:PPP}) one can write the probability of forming an
object of mass $A$ in the following form
\begin{equation}
P(A)\, =\, \frac{\partial P(A_i(t) < A)}{\partial A}\, =\,
\frac{m^{1/\alpha}}{1+1/\alpha}\cdot A^{-(1+1/\alpha)}\, .
\label{eq:PRES} 
\end{equation}
It means that we obtain in this way the following time-independent 
power-like distribution of abundance of elements,
\begin{equation}
P(A)\, \propto \, A^{-\gamma}\qquad {\rm with}\qquad \gamma = 1 +
\frac{1}{\alpha} = 1 + \frac{1}{nL\sigma} \, . \label{eq:FINP}
\end{equation}
In our case $\gamma = 7.5$ corresponds to $\alpha = nL\sigma =
0.154$ and this, for the tipical value of the cross section entering
here, $\sigma = 30$ mb, gives the thickness of the layer in which our
objects are produced equal to $nL = 5.1\cdot 10^{24}$ cm$^{-2}\, =
8.5$ g/cm$^2$. Assuming typical density on neutrons, $n\approx
10^{20}$ cm$^{-3}\, =1.6\cdot 10^{-4}$ g/cm$^3$ (and provided that
the so-called $r$-process nucleosynthesis conditions can be meet
\cite{R}) it corresponds to a rather thin layer of thickness equal to
$L=0.5$ km. 

\section*{Acknowledgments}
One of us (GW) would like to thank the Bogolyubow-Infeld Program
(JINR, Dubna) for financial help in attending the XXXII ISMD
conference where the above material has been presented.

\end{document}